\pdfoutput=1
\documentclass{appolb}
	
\usepackage{amsmath}
\usepackage{amssymb,bbm}
\usepackage{graphicx}
\usepackage{xcolor}
\usepackage{cite}
\usepackage{url}

\newcommand{\Pbbm}{\mathbbm{P}}
\newcommand{\Bbbm}{\mathbbm{B}}
\newcommand{\Pop}{\overleftarrow{\Pbbm}}
\newcommand{\Bop}{\overleftarrow{\Bbbm}}
\newcommand{\onebbm}{\mathbbm{1}}
\newcommand{\alfah}{{\hat{\alpha}}}
\newcommand{\betah}{{\hat{\beta}}}
\newcommand{\tbet}{{\tilde{\beta}}}

\begin{document}

%=============================================================================
%=============================================================================
%=============================================================================

\title{
%%%%%%%%%%%%%%%%%%
\vspace{-20mm}
\begin{flushright} \bf IFJPAN-IV-2011-9\\ \end{flushright}
\vspace{5mm}
%%%%%%%%%%%%%%%%%%
Fully NLO Parton Shower in QCD
\thanks
{This work is partly supported by the Polish Ministry of Science and Higher Education 
grant   No.\ 1289/B/H03/2009/37,
% by the Narodowe Centrum Bada\'n i Rozwoju
by the National Center for Research and Development
Grant LIDER/02/35/L-2/10/NCBiR/2011 and by the U.S.\ Department of Energy
under grant DE-FG02-04ER41299 and the Lightner-Sams Foundation. 
\hfill \\
  Presented by M. Skrzypek at the
XXXV International Conference of Theoretical
Physics ``Matter to the Deepest'', Ustro\'n, Poland, September 12--18, 2011.}
\author{M. Skrzypek,
 S. Jadach
\address{Institute of Nuclear Physics PAN,\\
ul. Radzikowskiego 152, 31-342 Krak\'ow, Poland }
\and
A. Kusina
\address{Department of Physics, Southern Methodist University,\\
Dallas, TX 75275, USA}
\and
W. P{\l}aczek
\address{
  Marian Smoluchowski Institute of Physics, Jagiellonian University,\\
  ul. Reymonta 4, 30-059 Krak\'ow, Poland}
\and
 M. Slawinska and O. Gituliar
\address{Institute of Nuclear Physics PAN,\\
ul. Radzikowskiego 152, 31-342 Krak\'ow, Poland 
}}
}

\maketitle

\begin{abstract}
{\em Abstract:}
The project of constructing a complete NLO-level Parton Shower Monte Carlo for
the QCD processes developed in IFJ PAN in Krak\'ow is reviewed. Four issues are
discussed: (1) the extension of the standard inclusive collinear factorization
into a new, fully exclusive scheme; (2) reconstruction  of the LO Parton Shower
in the new scheme; (3) inclusion of the exclusive NLO corrections into the hard
process and (4) inclusion of the exclusive NLO corrections into the evolution
(ladder) part. 

%\vspace{3mm}
%\centerline{\em Submitted To Acta Physica Polonica B}
\end{abstract}


\section{Introduction}

Precise calculations of the QCD cross-sections for the LHC are based on the
factorization procedure. It allows to divide the entire collision process into
separate parts: (1) the non-perturbative initial distribution of partons in the
incoming protons, to be taken from experiments in form of the Parton Density
Functions (PDFs); (2) the actual Parton Showers (PSs), both in the initial and
final states, described by the evolution equations resulting from the resummed
perturbative calculations; (3) the hard scattering calculated perturbatively to
a fixed order and (4) the non-perturbative hadronization of partons, described
by phenomenological models. There are a few approaches to the factorization,
amongst which the collinear factorization, well founded in the field theory, is
the most popular one. Let us recall a few milestones in its development. The
first, LO, resummations, dated to the early 1970s, are due to Gribov, Lipatov,
Altareli, Parisi and Dokshitzer \cite{DGLAP}. The general theorem has been
addressed in the papers of Ellis, Georgi, Machacek, Politzer
and Ross \cite{Ellis:1978sf} with the help of the axial gauge in which the
appealing ladder-like structure of the singularities has been proven.
The practical calculations of the NLO kernels were done by Floratos,
Ross and Sachrajda
\cite{Floratos:1977au,Floratos:1978ny} and then by Kalinowski, Konishi and
others \cite{Konishi:1979cb,Kalinowski:1980ju}, based on the operator product
expansion as well as by
Curci, Furmanski and Petronzio  \cite{Curci:1980uw,Furmanski:1980cm}
in diagrammatic approach in dimensionally regularized MS scheme.
Further reformulations of the
factorization scheme were due to Collins, Soper, Sterman \cite{Collins:1984kg},
Bodwin \cite{Bodwin:1984hc} and others in 1980s. Finally, twenty years later,
the NNLO kernels have been calculated by Moch, Vermaseren and Vogt
\cite{Moch:2004pa,Vogt:2004mw}. All the above papers concern the analytical
results. The Monte Carlo (MC) implementations, i.e.\ the PS programs, used in
the actual data analysis were developed in a different pace. The first
implementations: PYTHIA \cite{Sjostrand:1985xi} and HERWIG 
\cite{Marchesini:1988cf}, based on the (improved) LO calculations, were created
in mid-1980s and systematically developed since then. The next step -- the
complete NLO-based simulations have not been constructed yet, despite the fact
that the analytical results are known for more than 30 years! There are two
half-way solutions: MC@NLO  
\cite{Frixione:2006he} 
and POWHEG \cite{Nason:2004rx} which combine the NLO hard matrix element with
the LO-type PS (LO evolution). 
So far a LO accuracy of the MC programs was satisfactory in the data analysis,
as compared to the experimental precision of the hadronic experiments.  However,
with the start-up  of the LHC the situation has changed. The generic expected
experimental precision of LHC results is of the order of 1\%, and to match this
precision the MC simulations must be of the NLO type at least. 

The goal of the KRKMC project \cite{Jadach:2009gm,Jadach:2010aa} developed in
Krak\'ow is to fill in precisely this gap and to construct for the first time
the complete NLO-level PS program for QCD. 
In the following, after a short introduction to the collinear factorization, we
will briefly describe the four main ingredients of the KRKMC project: the
extended, exclusive collinear factorization scheme, the new LO PS scheme, 
exclusive NLO corrections to the hard process and exclusive corrections to the
ladder part. 

In the results presented here some simplifications are temporarily present:
only non-singlet kernels are included in the evolution, 
only $q\bar{q}\to W/Z$ hard process is included and 
non-running $\alpha_S$ is used.

\section{Extended exclusive collinear factorization scheme}
%%%%%%%%%%%%%%%%%%%%%%%%%%%%%%%%%%%%%%%%%%%%%%%%%%%%%%%%%%%%

\begin{figure}
\begin{center}
\includegraphics[width=8cm, height=3.2cm]{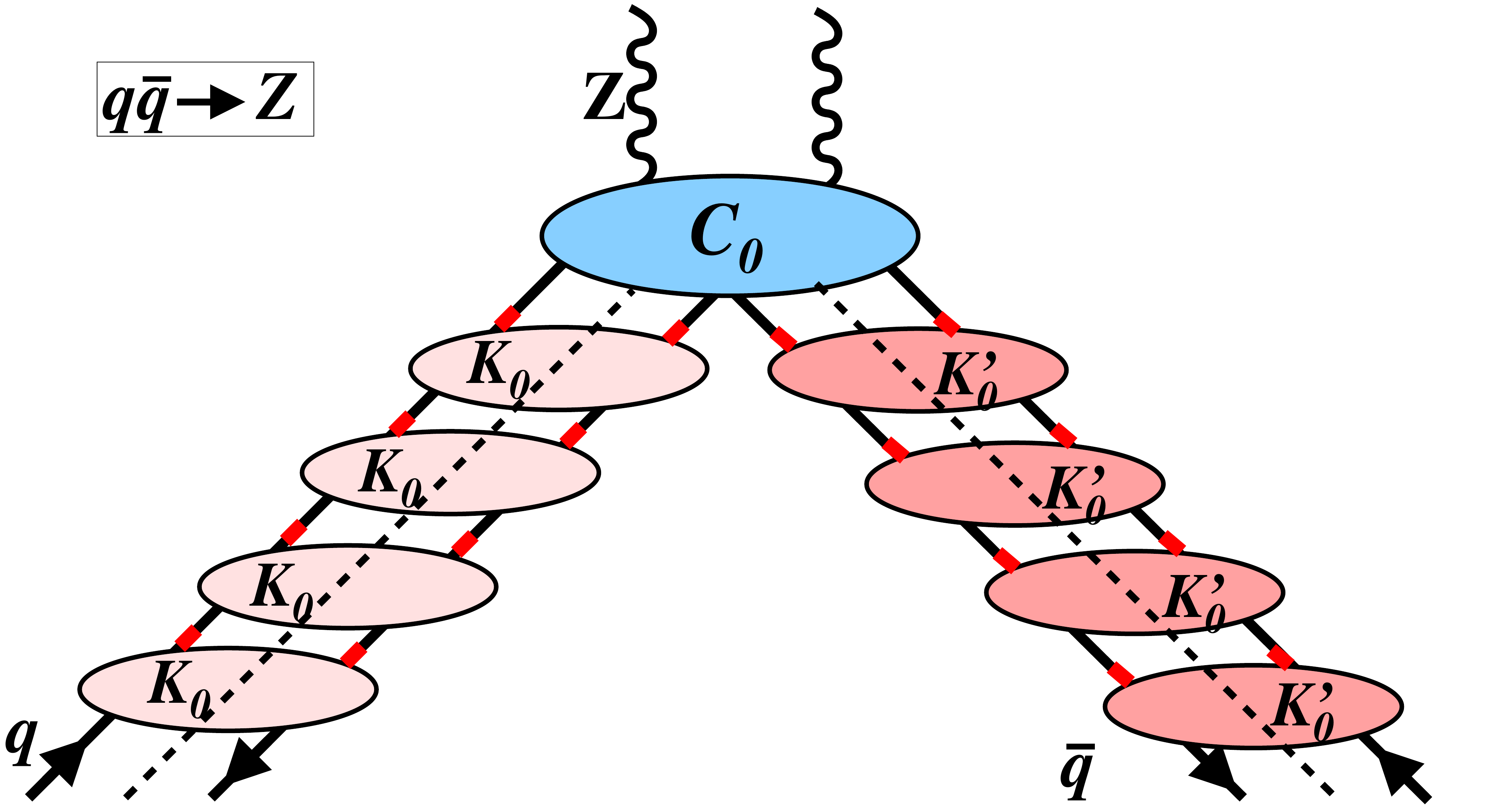}
\end{center}
\caption{The collinear factorization for the Drell-Yan type process}
\label{drelly}
\end{figure}

The standard collinear factorization rearranges the squared matrix element into
a sum of two-particle-irreducible universal kernels $K_0$, and a 
process-dependent function $C_0$, as shown in Fig.\ \ref{drelly} for the 
Drell-Yan type process (use of the physical gauge is instrumental here!).  
The collinear singularities are located in the $K_0$ functions. At this point
one introduces the projection operator $\Pbbm$ which separates the singular part
$\Pbbm K_0$ from the non-singular rest $(\onebbm-\Pbbm) K_0$, leading to the
rearrangement 
\begin{align}
 &|M|^2=C_0\cdot \frac{1}{1-K_0}
 = C\otimes(1+K+K\otimes K +K\otimes K\otimes K+...),
\label{eqM}
\\
&C=  C_0\cdot\frac{1}{1-(1-\Pbbm)\cdot K_0},\;\;\;\;
K=\Pbbm K_0\cdot\frac{1}{1-(1-\Pbbm)\cdot K_0}
\end{align}
where the reorganized kernel and the hard process, $K$ and $C$, are to be
calculated up to a requested accuracy: LO, NLO, etc. The dot $(\cdot)$ denotes
four-dimensional integration, whereas the $\otimes$ is a one-dimensional
integral (convolution) over the longitudinal momentum (lightcone) variable
$x^+$. The transverse momentum degrees of freedom are integrated out. 

Why is the above scheme not good for precision MC simulation? The three main
reasons are: 
(1) It does not conserve four momenta. In order to reduce $\cdot$ into 
$\otimes$, the $\Pbbm$ operator allows for unphysical configurations of
momenta. 
(2) There are very strong over-subtractions (cancellations) present. Eq.\
(\ref{eqM}) is a geometrical series whereas we expect the final result to be of
the exponential form. 
(3) The scheme is defined in dimensional regularization (pole-part extraction),
whereas the MC simulation must be done in four dimensions. 
How can one cure these drawbacks? (1) The projection operator $\Pbbm$ has to be
redefined. (2) The time-ordered exponential must follow directly from the
construction of the modified factorization expression 
(\ref{eqM}). (3) The geometrical regularization has to be introduced for the
real emissions instead of the dimensional one. 

We define a new $\Pop'$ operator and also introduce a finite-part operator 
$\Bbbm$. To the lowest order we have $\Bop_\mu [ K_0 ]  = K_0 - \Pop'_\mu\{
K_0 \}$, as expected. In higher orders $\Bop$ is defined recursively: 
$
\Bop_\mu [ K_0 \cdot K_0 ]\!  =\!
K_0 \cdot K_0 \!
 -\Pop'_\mu\{^{s_2} K_0 \}\! \cdot \Pop'_{s_2}\{^{s_1} K_0 \}\!
 -\Pop'_\mu\{^{s_2} K_0 \!   \cdot \Bop_{s_2}[K_0] \}\!
 -\Bop_\mu[ K_0 ]\!           \cdot \Pop'_{\mu}\{ K_0 \},
$
and so on.

$\Pop'_\mu$ sets the upper limit $\mu$ on the phase space for all
real partons  {towards the hadron} 
using the kinematical variable $s(k_1,..,k_n)<\mu$ 
where
$s$ stands for virtuality,
maximal rapidity, maximal $k^T$, etc.
The notation { $\{^{s_i} A \}$} defines
$s=s_i$,
e.g.\
$
   \Pop'_\mu\{^{s_3} A \}\cdot
   \Pop'_{s_3}\{^{s_2} A \}\cdot\Pop'_{s_2}\{^{s_1} A \}
$ 
means 
$
\theta_{\mu>s_3>s_2>s_1}
$
instead of
$
   \Pop_\mu\{^{s_3} A \}\cdot
   \Pop_{\mu}\{^{s_2} A \}\cdot\Pop_{\mu}\{^{s_1} A \}
$ 
corresponding to
$
\theta_{\mu>s_3}\; \theta_{\mu>s_2}\; \theta_{\mu >s_1}
$ (CFP-like).
Also, 
$\Pop'_\mu(A)$ extracts a singular part from an integrand $A$
(not from the integral $\int A$ like CFP!).

An exclusive parton density function is then defined as the integrand in:
\begin{align}
D(\mu)=
\exp_{TO}\left(\Pop'_\mu
\left\{^s\; K_0 \cdot
\Bop_s\; \left[ \frac{1}{1-K_0} \right]
\right\}
\right),\;\;\partial_\mu D(\mu,x) = P\otimes D(\mu) (x),
\label{DDD}
\end{align}
and at the inclusive level it fulfills the standard DGLAP equation.

\section{LO Parton Shower revisited}
%%%%%%%%%%%%%%%%%%%%%%%%%%%%%%%%%%%%%

Having established the general factorization framework, we now turn to the LO
MC. It is the basis of the project. All the NLO effects will be added on top of
it by means of the rejection techniques. The strategy is standard -- one 
combines twice the formula (\ref{DDD}) (a Drell--Yan case): 
\begin{align}
\sigma(C_0^{(0)}  \Gamma_F^{(1)} \Gamma_B^{(1)} )=
\sum_{n_1=1}^\infty
\sum_{n_2=1}^\infty
\big\{\sigma\big[
  C_0^{(0)} 
  (\Pop'  K_{0F}^{(1)})^{n_1}
  (\Pop'' K_{0B}^{(1)})^{n_2}
\big]\big\}_{T.O.}
\label{LOLO}
\end{align}
and parametrizes it in terms of the Sudakov variables ($\alpha_i,\beta_i$ for
emitted partons and $x_i$ for virtual, ``ladder", ones) that can be directly
generated. There are two technical problems to be resolved: (1) the constraint
on the value of the final $x_{F/B}=\prod  z_{F/B}^i$ (necessary for resonant
processes) and (2) complete coverage of the phase space without any gaps. 
The standard solution of the problem (1) is the ``backward evolution" 
\cite{Sjostrand:1985xi}, which uses a pretabulated grid of PDFs. In order to
avoid this complication we proposed a different algorithm which imposes the 
constraint on top of the normal ``forward" evolution, see  
\cite{Jadach:2005bf,Jadach:2005yq} for details.  
The problem (2) is solved by a kinematical mapping of the original phase space
to the ``tangent space"  ($k_i,\alpha_i,\beta_i\to\bar{k_i},\alfah_i,\betah_i$).
The mapping we proposed in Ref.\ \cite{Jadach:2011cr} is a plain rescaling,
although defined in a recursive way: 
\begin{align}
{k_{\pi_i}=\lambda_{i} \bar{k}_{\pi_i},\quad}
\lambda_{i}= 
\frac{s(\bar{x}_{i-1}-\bar{x}_{i})}%
     {2(P-\sum_{j=1}^{i-1} k_{\pi_j} )\cdot \bar{k}_{\pi_i}},\quad
i=1,2,...,n_1+n_2.
\end{align}
The rescaling factors $\lambda_{i}$ are chosen such that 
\begin{align}
\bar{s}_i= s\bar{x}_i
=s\prod\limits_{\pi_j\in F}^i \hat{z}_{F\pi_j}
  \prod\limits_{\pi_j\in B}^i \hat{z}_{B\pi_j}
=(P-\sum\limits_{j=1}^{i} k_{\pi_j})^2
=(P-\sum\limits_{j=1}^{i} \lambda_j \bar{k}_{\pi_j})^2.
\end{align}
This mapping has three important features:
(1) it { preserves angles}, i.e.\ rapidity ordering, and the upper limit of the
phase-space (integration) in rapidity; 
(2) it { preserves soft factors} ($d\alpha/\alpha\dots$), i.e.\ the original
behavior in the soft limit; 
(3) it covers the phase-space completely, without any gaps. 

The complete LO MC algorithm looks as follows:
\begin{enumerate}
\item
the variables $\hat{z}_F$ and $\hat{z}_B$ are generated by the FOAM MC Sampler
\cite{foam:1999,foam:2002},
\item
the four-momenta $\bar{k}_i^\mu$ are generated separately
in the F and B parts of the phase space with the constraints
$\sum_{j\in F} \alfah_j=1-\hat{z}_F$
and $\sum_{j\in B} \betah_j=1-\hat{z}_B$,
\item
the double-ordering permutation $\pi$ is established,
\item
the rescaling parameter $\lambda_1$
is calculated; $k_{\pi_1}=\lambda_1 \bar{k}_{\pi_1}$ is set, such that
$(P-k_{\pi_1})^2=sx_1$,
\item
the parameter $\lambda_2$ is calculated and
$k_{\pi_2}=\lambda_2 \bar{k}_{\pi_2}$ is set, such that
$(P-k_{\pi_1}-k_{\pi_2})^2=sx_2=sz_{\pi_1}z_{\pi_2}$ and so on,
\item
in the rest frame of $\hat{P}= P-\sum_j k_{\pi_j}$
four-momenta $q_1^\mu$ and $q_2^\mu$ are generated according to the Born
angular distribution.
\end{enumerate}
Exact analytical integration
of the LO MC distributions of eq.\ (\ref{LOLO}) over the multigluon phase space
is possible (we use the rapidity ordering with $\Xi$ being the rapidity of the
produced boson, or equivalently the splitting point between the forward and
backward hemispheres):\\ 
\[
\sigma(C_0^{(0)}  \Gamma_F^{(1)} \Gamma_B^{(1)} )=
\int_0^1 d\hat{x}_F\;d\hat{x}_B\;
D_F(\Xi, \hat{x}_F)\;D_B(\Xi, \hat{x}_B)\;
\sigma_B(s\hat{x}_F\hat{x}_B),
\]
with two PDFs obeying the DGLAP non-singlet LO evolution equation:
\[
 \frac{\partial }{ \partial \Xi} D_F(\Xi,x) = [P \otimes D_F(\Xi)](x).
\]

\section{Exclusive NLO corrections to the hard process}
%%%%%%%%%%%%%%%%%%%%%%%%%%%%%%%%%%%%%%%%%%%%%%%%%%%%%%%

Having described the ``underlying" LO MC, we proceed now with the NLO
corrections. We begin with the hard process part. The NLO correction consists
of a one-parton real emission and the matching first order virtual correction.
It is included by means of a MC weight. In the case of the Drell--Yan
process (Fig.\ \ref{drelly}) the weight reads: 
\begin{align}
W^{NLO}_{MC} \! =
1+\Delta_{S+V}
+\!\sum_{j\in F} 
 \frac{\tbet_1(\hat{s},\hat{p}_F,\hat{p}_B;a_j, z_{Fj})}%
      {\bar{P}(z_{Fj})\;d\sigma_B(\hat{s},\hat\theta)/d\Omega}
+\!\sum_{j\in B} 
 \frac{\tbet_1(\hat{s},\hat{p}_F,\hat{p}_B;a_j, z_{Bj})}%
      {\bar{P}(z_{Bj})\;d\sigma_B(\hat{s},\hat\theta)/d\Omega}.
\nonumber
\end{align}
The {IR/colinear-finite} { real} emission part $\tbet_1$ (numerators) is a
fully-diffe\-rent\-ial real emission matrix element minus two counter-terms
(the counter-terms are LO collinear distributions generalized to off-collinear
regions): 
\begin{align}
\begin{split}
&
\tbet_1(\hat{p}_F,\hat{p}_B;q_1,q_2,k)=
\Big[
  \frac{(1-\alpha)^2}{2}
  \frac{d\sigma_{B}}{d\Omega_q}(\hat{s},\theta_{F1})
 +\frac{(1-\beta)^2}{2}
  \frac{d\sigma_{B}}{d\Omega_q}(\hat{s},\theta_{B2})
\Big]
\\&~~
-\theta_{\alpha>\beta}
 \frac{1+(1-\alpha-\beta)^2}{2}
 \frac{d\sigma_{B}}{d\Omega_q}(\hat{s},\hat\theta)
-\theta_{\alpha<\beta}
 \frac{1+(1-\alpha-\beta)^2}{2}
 \frac{d\sigma_{B}}{d\Omega_q}(\hat{s},\hat\theta).
\end{split}
\nonumber
\end{align}
The denominators are just the LO ``underlying" differential distributions:
numerators of the DGLAP LO kernels $\bar{P}(z_{Fj/Bj})$ times the  Born cross
section $d\sigma_B(\hat{s},\hat\theta)/d\Omega$. 
The sums go over all emitted partons in both the F and B hemispheres. This way
all partons contribute to the hard scattering and there is no problem of
defining the ``last", ``hardest", etc.\ one. 

Similarly to the real emission part, the 
{virtual+soft} correction is defined as a difference between the complete
inclusive result and two inclusive MC-type counterterms 
\[
\Delta_{V+S}=\Bigl({  D_{DY}^{{MS}}(z)} 
 -{2C_{ct}^{MC}(z)}\Bigr)\Big|_{\delta\hbox{\scriptsize -part}}
=
 \frac{C_F \alpha_s}{\pi}\;\left( \frac{2}{3}\pi^2 -\frac{5}{4} \right).
\]
$D_{DY}^{{MS}}(z)$ can be taken from the literature (eq.\ (89) in Ref.\
\cite{Altarelli:1979ub}) or re-calculated from the Feynman graphs
\cite{gituliarinprep}. Details, in particular on the construction of the
counterterms, can be found in \cite{Jadach:2011cr}. Note that the $
\Delta_{V+S}$ is {kinematics independent} (it is a constant). 

As in the LO case, the exact analytical integration 
of the NLO MC distributions over the multigluon phase space
is possible, and leads to: 
\[
\begin{split}
&\sigma(C_0^{(1)}  \Gamma_F \Gamma_B )=
\int\limits_0^1 d\hat{x}_F\;d\hat{x}_B\; dz\;
D_F(\Xi, \hat{x}_F)\;D_B(\Xi, \hat{x}_B)\;
\sigma_B(sz\hat{x}_F\hat{x}_B)
\\&~~~~~~~~~~~~~~~~~~~~~~~~~~~~~~\times
\bigg\{
\delta_{z=1}(1+\Delta_{S+V})
+{C^{MC}_{2r}(z)}
\bigg\},
\end{split}
\]
where
\begin{equation}
{C^{MC}_{2r}(z)} 
= - \frac{C_F \alpha_s}{\pi}(1-z).
\label{mcpres}
\end{equation}
The above inclusive NLO correction to the hard scattering in the MC scheme
differs from the MS correction (eq.~(90) in \cite{Altarelli:1979ub})
\begin{equation}
{C^{{MS}}_{2r}(z)}
=\frac{C_F \alpha_s}{\pi}
 \frac{\bar{P}(x)}{1-z}
 [4\ln(1-z) -2\ln z].
\label{msbar}
\end{equation}
As we see, the MC correction in eq.\ (\ref{mcpres}) is a simple, regular
polynomial,
whereas the MS one in eq.\ (\ref{msbar}) contains the singular logarithmic
terms
$\ln(1-z)/(1-z)$. These terms originate from the ``mistreatment" of the phase
space done by the $k_T$ ordering implied by MS, see also 
\cite{Altarelli:1979ub}.

To summarize, the new NLO MC factorization scheme has the following interesting
features: 
 (1)
The NLO corrections are added on top of the LO MC with a simple, positive weight.
 (2)
There is no need to correct for the difference
in the collinear counter-terms of the PSMC and MS schemes provided PDFs are
in the PSMC scheme. 
 (3)
The virtual+soft corrections are completely kinematics independent
-- all the complicated $d\Sigma^{c\pm}$
contributions of the MC@NLO scheme are absent.
 (4)
There is a built-in resummation of the $\frac{\ln^n(1-x)}{1-x}$ terms
(demonstrated for $n=1$).

\section{Exclusive NLO corrections to the ladders}
%%%%%%%%%%%%%%%%%%%%%%%%%%%%%%%%%%%%%%%%%%%%%%%%%%%

The NLO corrections to the multi-emission ladder part have not been included in
any of the existing PSs yet, except for a partial proposal of Ref.\
\cite{Tanaka:2011ig}. The scheme presented here is the first  complete
solution for the non-singlet evolution. Our scheme is based on the reweighting
technique. Once again, we begin with the LO ``underlying" differential
distribution for the single ladder: 
\begin{displaymath}
\begin{split}
\bar{D}^{LO}(x,Q)=& \sum\limits_{n=0}^\infty
\raisebox{-30pt}{\includegraphics[height=30mm]{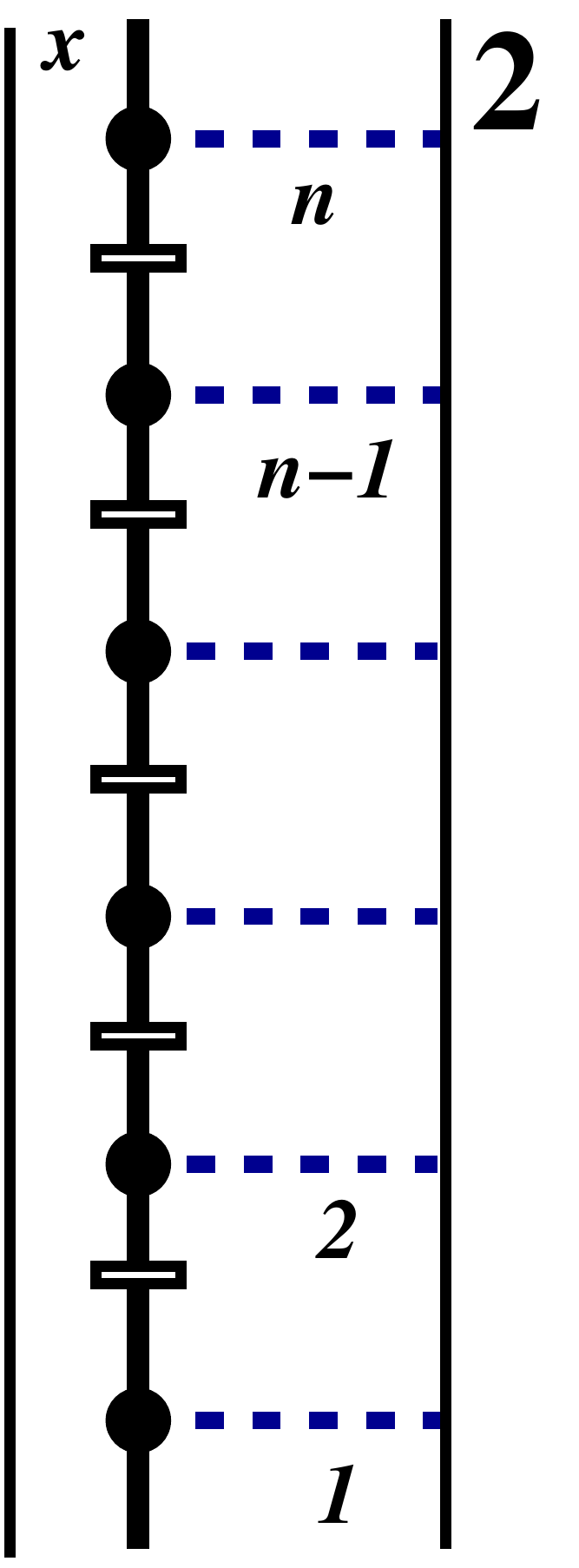}}\!
 = e^{-S}
\sum\limits_{n=0}^\infty
\prod\limits_{i=1}^n
\frac{ d^3 k_i}{k_i^0}
\theta_{Q>a_i>a_{i-1}}
\rho^{(0)}_{1}(k_i),
\\
\rho^{(0)}_{1}(k_i)
=& \frac{2C_F^2\alpha_s}{\pi} \frac{1}{k_i^{T2}} \frac{1+z^2}{2}.
\end{split}
\end{displaymath}
The NLO corrected distribution looks as follows:
\begin{displaymath}
\begin{split}
&\bar{D}^{NLO}(x,Q)=
\\
&=
e^{-S}
\sum\limits_{n=0}^{\infty}
\Bigg\{
\raisebox{-30pt}{\includegraphics[height=30mm]{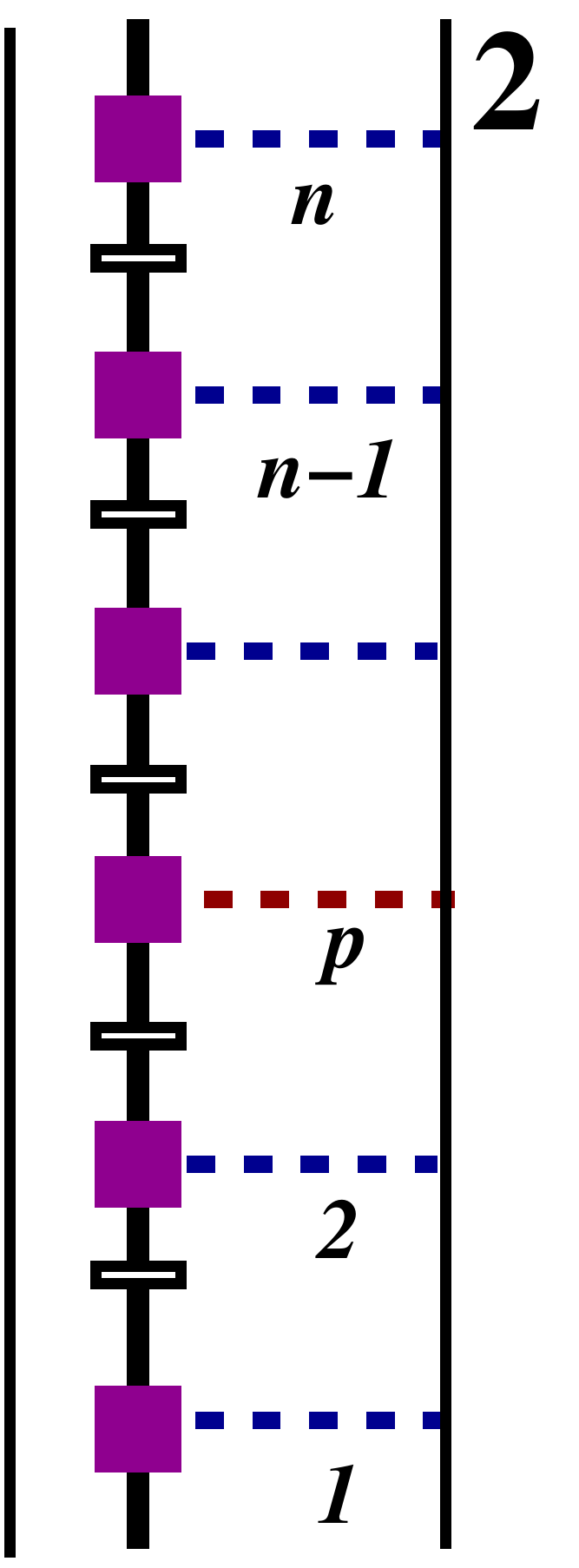}}
\!\! +\!
\sum\limits_{p_1=1}^{n}
\sum\limits_{j_1=1}^{p_1-1}
\raisebox{-30pt}{\includegraphics[height=30mm]{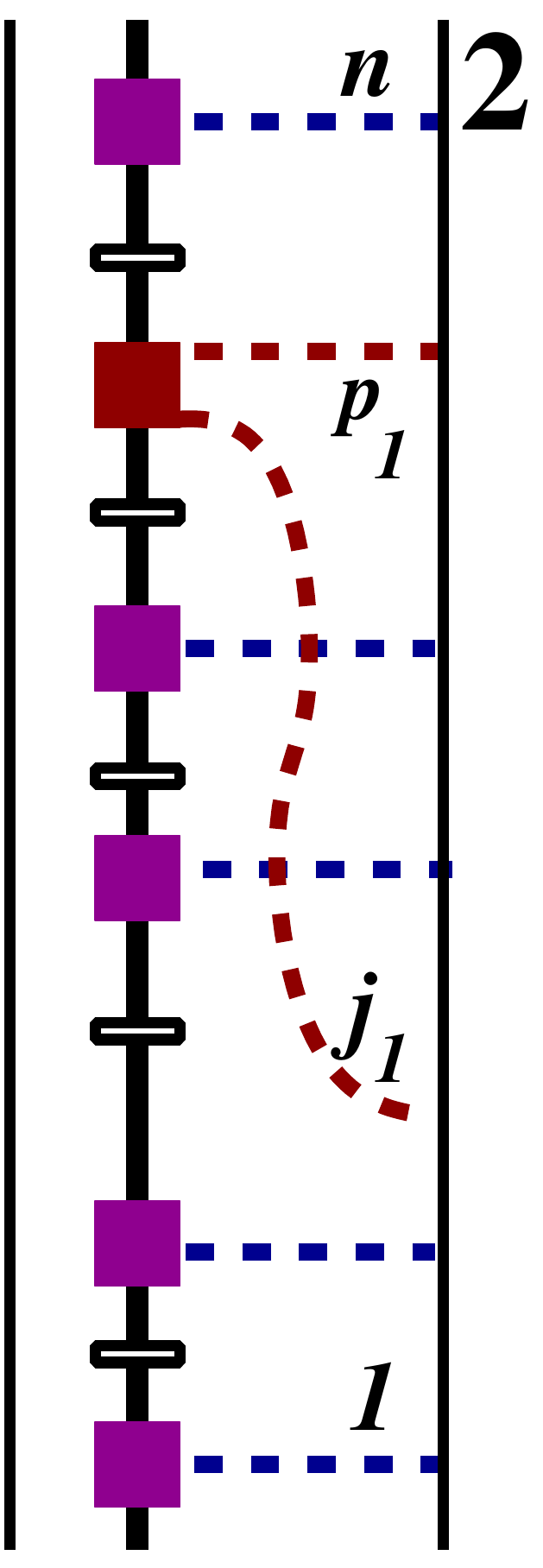}}
 \!+\!
\sum\limits_{p_1=1}^{n}
\sum\limits_{p_2=1}^{p_1-1}
\sum\limits_{j_1=1 \atop j_1\neq p_2}^{p_1-1}
\sum\limits_{j_2=1 \atop j_2\neq p_1,j_2}^{p_2-1}
\raisebox{-30pt}{\includegraphics[height=30mm]{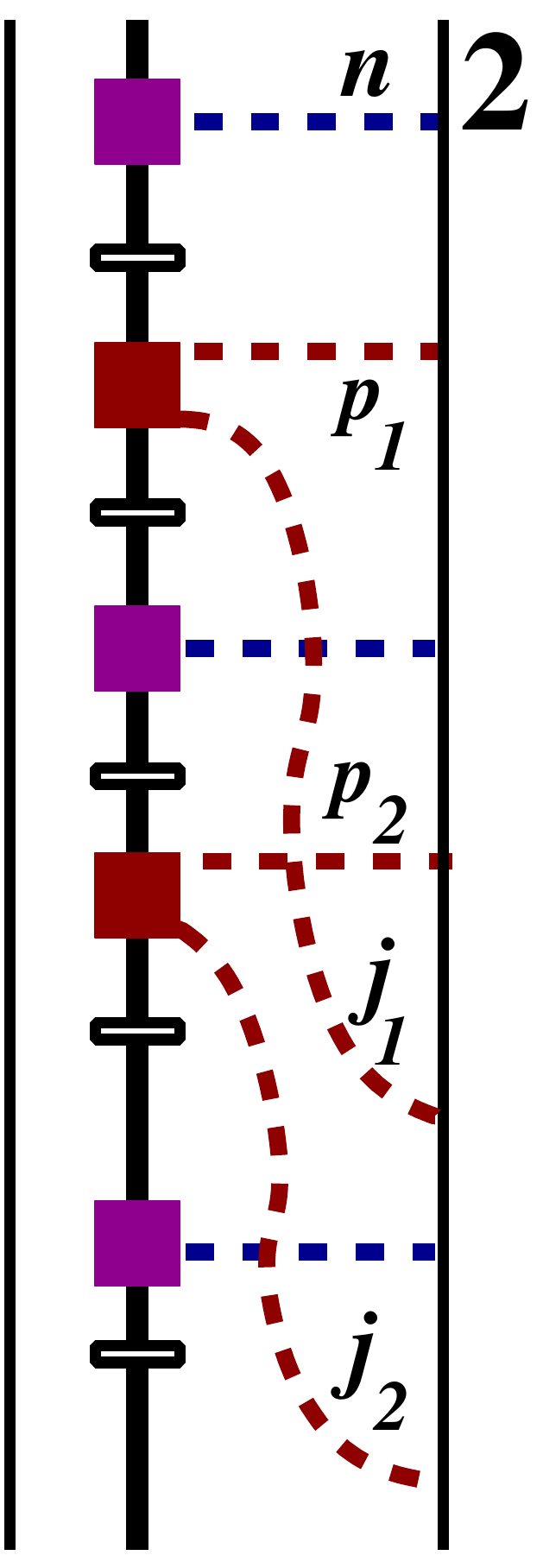}}
+\dots\Bigg\}
\\&
= e^{-S}
\Bigg\{ \delta_{x=1}
+\sum_{n=1}^\infty\;
\bigg( \prod_{i=1}^n\; 
    \int\limits_{Q>a_i>a_{i-1}}\!\!\!\!\!\! 
    \frac{ d^3 k_i}{k_i^0}\;
    \rho^{(0)}_{1}(k_i){ \beta_0^{(1)}(z_i)}
\bigg)
\delta_{x=\prod_{j=1}^n x_j}
\bigg[
1 + 
\\ &
 + \sum\limits_{p=1}^{n} \sum_{j=1}^{p-1}
    { W(\tilde{k}_p, \tilde{k}_j)}+
 \sum\limits_{p_1=1}^{n}
\sum\limits_{p_2=1}^{p_1-1}
\sum\limits_{j_1=1 \atop j_1\neq p_2}^{p_1-1}
\sum\limits_{j_2=1 \atop j_2\neq p_1,j_2}^{p_2-1}
\!\!\!\!
{ W(\tilde{k}_{p_1}, \tilde{k}_{j_1})}
{ W(\tilde{k}_{p_2}, \tilde{k}_{j_2})}
+\dots
\bigg]
\Bigg\}.
\end{split}
\end{displaymath}
The above formula might look complicated, but in fact its structure is simple.
The LO ladder is multiplied by appropriate non-singular NLO weights: in the
first term by the virtual weights $\beta_0^{(1)}$ (in the picture dots are
replaced by squares); 
in the second term one NLO real correction, $W(\tilde{k}_p, \tilde{k}_j)$, is
added with all possible choices of the first and second momentum (two sums are
visible); the third contribution is identical to the second one but with two
NLO corrections and so on. 
The actual definitions of the weights are intuitively obvious -- they are
ratios of the NLO to LO distributions (with counterterms if needed) and can be
graphically represented as: 
\begin{displaymath}
{\beta_0^{(1)}}=
\frac{%
 \left| \raisebox{-8pt}{\includegraphics[height=8mm]{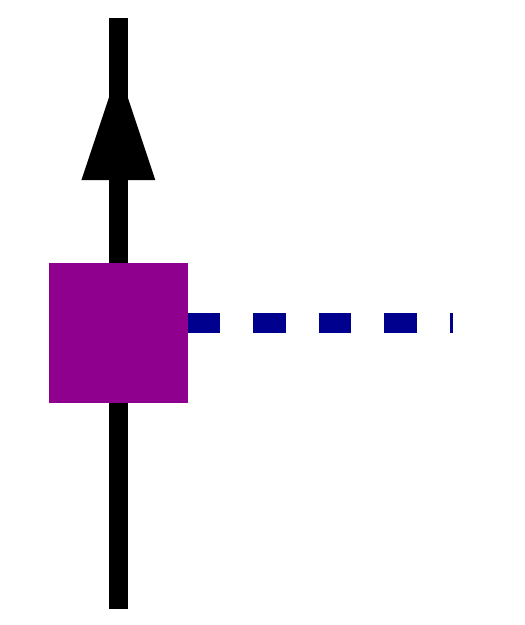}}
 \right|^2}%
{ \left|\raisebox{-8pt}{\includegraphics[height=8mm]{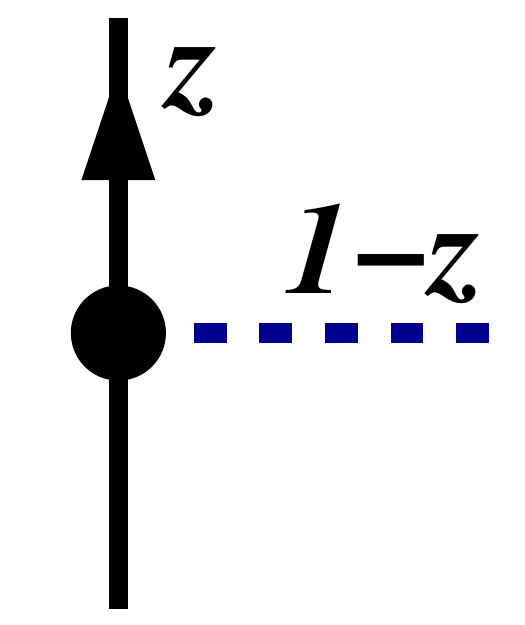}}
 \right|^2
} 
=
1+2\Re(\Delta_{_{ISR}}^{(1)}),
\;\;\;\;
 {W(k_2,k_1)}=
\frac{%
 \left| \raisebox{-8pt}{\includegraphics[height=8mm]{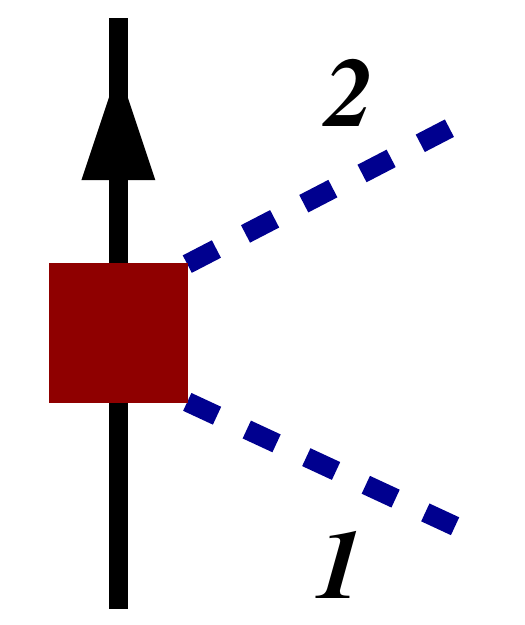}}
 \right|^2
}{%
 \left| \raisebox{-8pt}{\includegraphics[height=8mm]{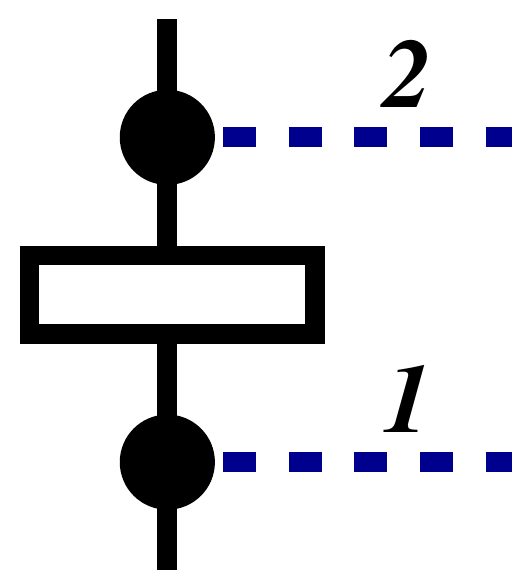}}
 \right|^2
}=
\frac{%
 \left| \raisebox{-8pt}{\includegraphics[height=8mm]{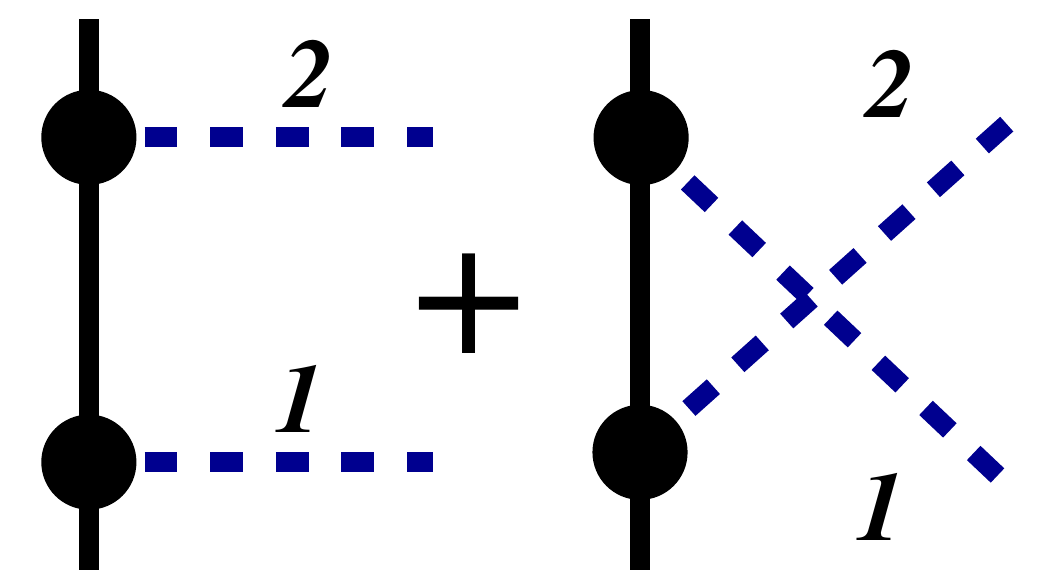}}
\right|^2
}{%
 \left| \raisebox{-8pt}{\includegraphics[height=8mm]{xBr2ReCt.pdf}}
 \right|^2
}\; -1.
\end{displaymath}
We have recalculated the real emission weight $W(k_2,k_1)$ in the new
factorization scheme \cite{Jadach:2011kc,Kusina:2011xh}  and we are in the
process of recalculating the virtual weight \cite{gituliarinprep}.
We have tested numerically the above algorithm by comparing the inclusive
distribution $\bar{D}^{NLO}(x,Q)$ from our exclusive MC with the standard
inclusive result. 
We obtained three-digit agreement limited by the statistics, see
\cite{Jadach:2009gm,Jadach:2010aa} for details.  

In the above we have ignored the contribution to $W(k_2,k_1)$ from the
gluon-pair production. This contribution has additional singularity in the
limit of a vanishing mass of the emitted gluon pair. In the inclusive approach
this singularity is cancelled by the corresponding virtual one. In the MC
exclusive simulation this is not possible and the singularity has to be
included into the ``underlying" distribution, otherwise it would ruin the
convergence of the algorithm. Fortunately, as it is well known, this
singularity is a part of the LO corrections to the final state and can be
included into the algorithm at the expense of a few additional combinatorial
sums over the final-state emissions. Graphically it looks as follows: 
\begin{displaymath}
e^{-S_{_{ISR}}-S_{_{FSR}}}
 \sum\limits_{n,m=0}^{\infty}
 \sum\limits_{r=1}^{m}
\raisebox{-20pt}{\includegraphics[height=20mm]{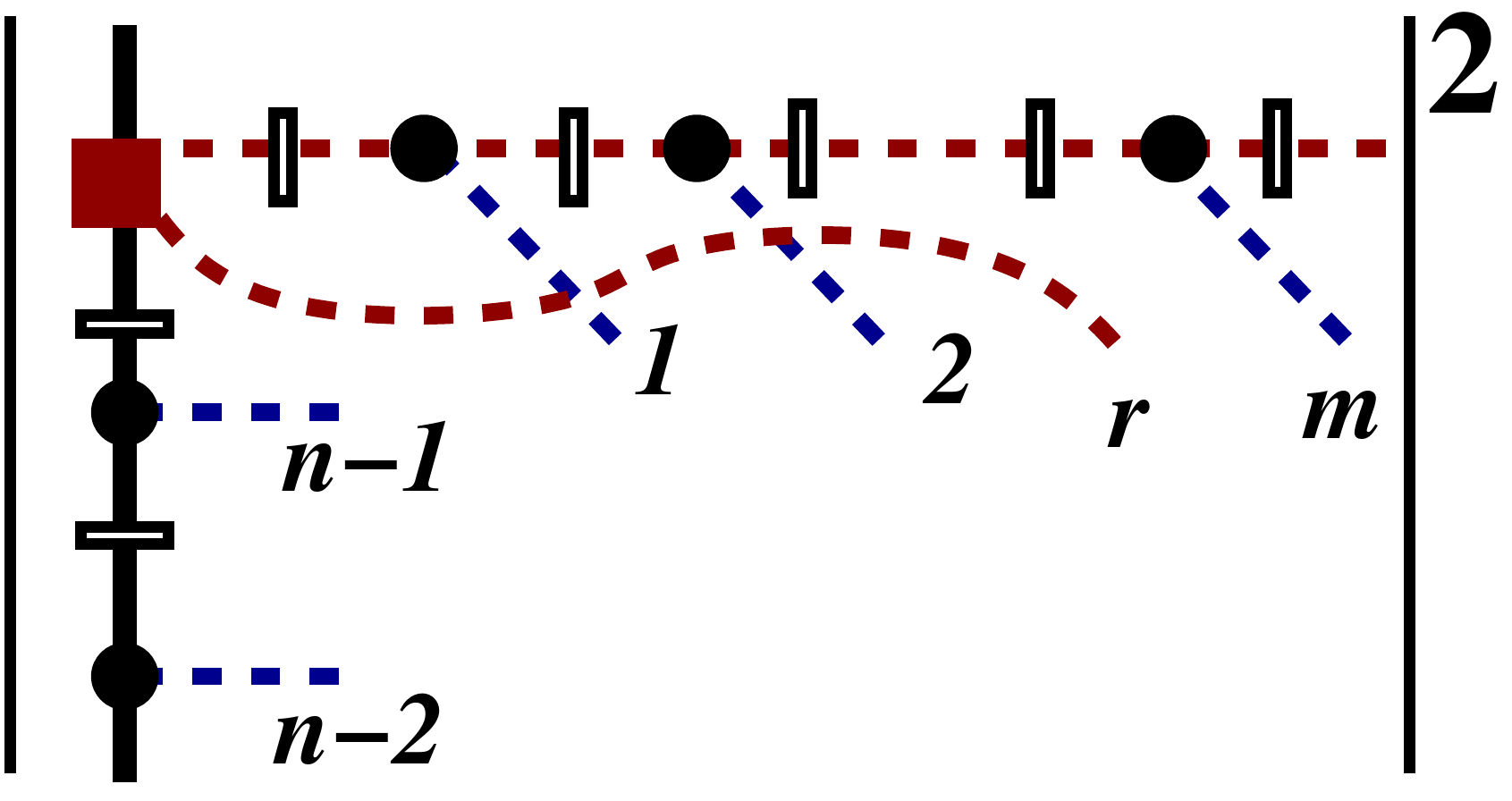}},
\end{displaymath}
where the Sudakov form-factor { $S_{_{FSR}}$} is subtracted in the virtual part:
\begin{displaymath}
\left|
 \raisebox{-9pt}{\includegraphics[height=9mm]{xBrBet0ISR.pdf}}
\right|^2
=\big(1+2\Re(\Delta_{_{ISR}}+V_{_{FSR}} {-S_{_{FSR}}})\big)
\left|
  \raisebox{-9pt}{\includegraphics[height=9mm]{xBrBorn.pdf}}
\right|^2,
\end{displaymath}
and the FSR real {counterterm} is subtracted together with the ISR one:
\begin{displaymath}
\left|
  \raisebox{-9pt}{\includegraphics[height=9mm]{xBrBetISR.pdf}}
\right|^2
=\left|
  \raisebox{-9pt}{\includegraphics[height=9mm]{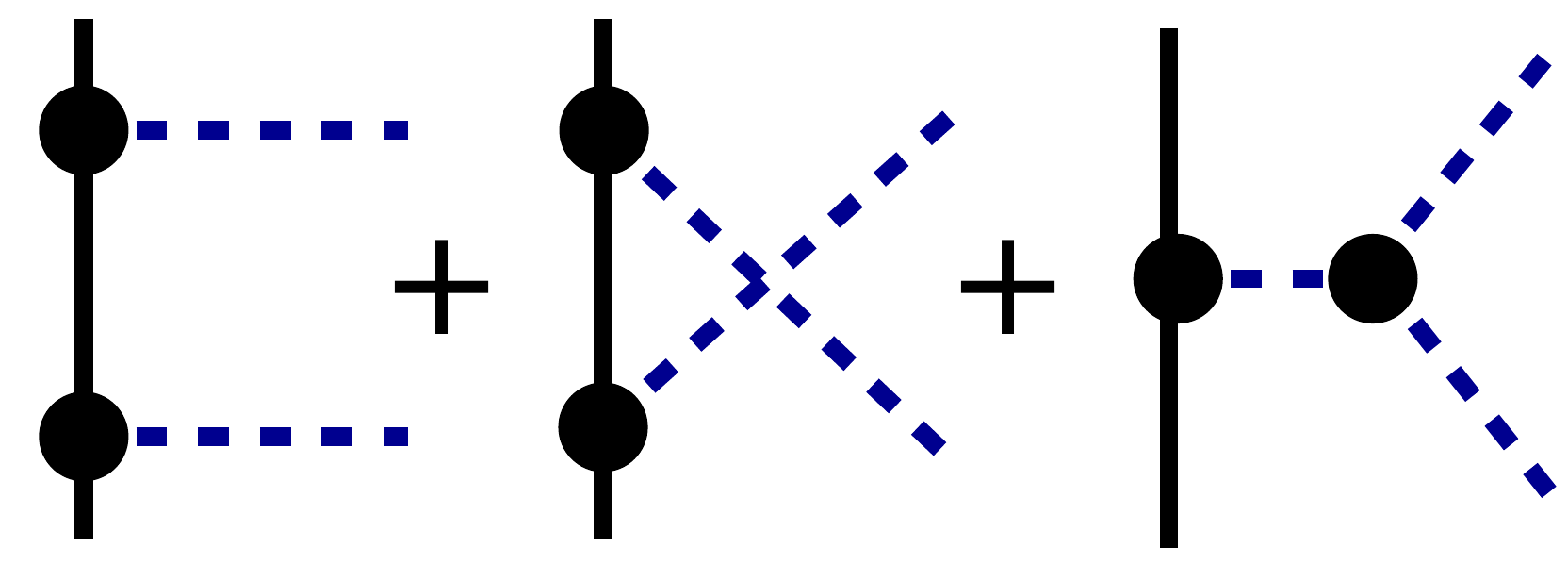}}
\right|^2
-\left|
  \raisebox{-9pt}{\includegraphics[height=9mm]{xBr2ReCt.pdf}}
\right|^2
{
-\left|
  \raisebox{-9pt}{\includegraphics[height=9mm]{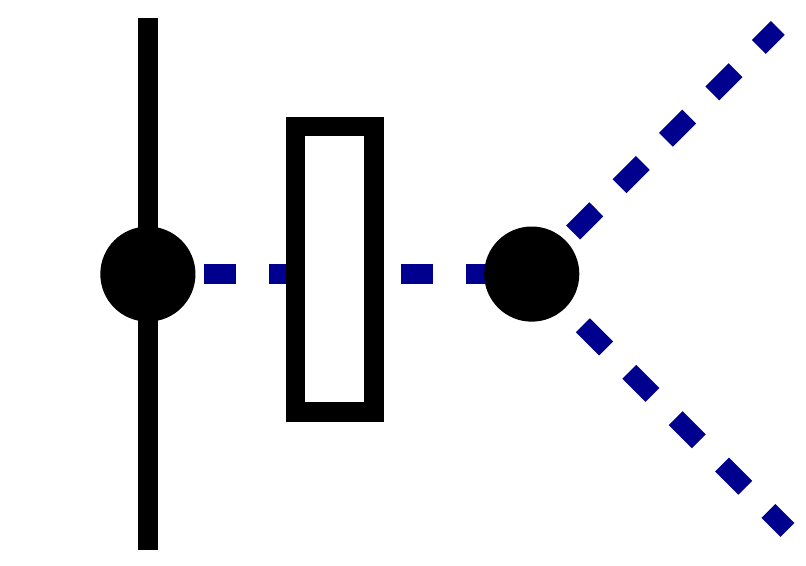}}
\right|^2}.
\end{displaymath}
Both the virtual and real corrections (weights) are now regular. All
singularities have cancelled separately in each weight, allowing for
construction of the efficient MC algorithm. The exact analytical integration is 
possible in this case and we have used this result to perform numerical tests
of the MC program with the percent-level precision.

\section{Summary}
We have briefly reviewed the most important results of the first ever, complete
NLO, fully exclusive, PS project under  development in Krak\'ow:
%\\
(1) Extension of the collinear factorization,
better suited for the MC implementation, is defined.
%\\
(2) LO PSMC is (re-)constructed from scratch,
in a way compatible with the new factorization scheme.
%\\
(3) The NLO parts of the hard process and the evolution kernels 
are recalculated in the new scheme 
(so far the non-singlet NLO exclusive kernels only).
%\\
(4) The differences between
the new MC and standard MS schemes are understood, 
keeping the universality (process independence) in mind.
%\\
(5) The proposed solution has advantages as compared to the other
techniques of adding the NLO corrections to the hard process and
it is completely new for the ladder parts.
%\\
(6) Implementation in the MC has been tested at the prototype level with the
relative precision of $10^{-3}$.

%=============================================================================
%=============================================================================
%=============================================================================

%%\bibliographystyle{utphys_spires}
%\bibliographystyle{h-physrev3}
%\bibliography{radcor}

%=============================================================================

%=============================================================================
%=============================================================================
\end{document}